\documentclass[a4]{article}
\usepackage[left=2cm,right=2cm,top=2cm,bottom=2cm,bindingoffset=0cm]{geometry}
\usepackage{multirow}

\usepackage{amsfonts, amsmath, amssymb}
\usepackage{graphicx, float}
\usepackage{subcaption}

\newcommand{\be}{\begin{equation}}
\newcommand{\ee}{\end{equation}}
\newcommand{\Tr}{{\rm Tr}}

\graphicspath{{./pictures/}}

\renewcommand{\Re}{\text{Re}}
\renewcommand{\Im}{\text{Im}}

\begin{document}
	\title{LOW-ENERGY EFFECTIVE LAGRANGIAN OF THE TWO-HIGGS-DOUBLET MODEL}
	\author{M.S.~Dmytriiev$^1$, V.V.~Skalozub$^2$, \\ $^1$dmytrijev\_m@ffeks.dnu.edu.ua, $^2$Skalozubv@daad-alumni.de, \\
		Oles Honchar Dnipro National University, \\
		72, Gagarin Ave., Dnipro 49010, Ukraine}
	
	\maketitle
	
	\begin{abstract}
		We consider a decoupling scenario within the two-Higgs-doublet model (2HDM) with small CP-violation. Mass eigenstates of this model include one neutral scalar field with the mass of the Standard model (SM) Higgs boson and four other scalars, which decouple at low energies.  We derive the effective operators of interactions between the SM fermions and the lightest scalar particle of the  model. The coefficients at these  operators are expressed in terms of  the 2HDM parameters. The scattering processes  affected by this effective Lagrangian are identified.
		
	\end{abstract}

    \section{Introduction}
    Nowadays, the Standard model is the best experimentally proven theoretical description of interactions between elementary particles. However, there are physical phenomenons which could not be explained within the SM, such as baryon asymmetry in the Universe, neutrino masses, dark matter, etc. To address these problems, many different models were proposed, which extend the SM with different new particles. Observable predictions of these models had been tested in experiments, but no new states beyond the SM were found so far. This could happen because of different reasons. In our paper, we consider the case when masses of new particles are much bigger than the collision energies used in the experiments. Hence, their contributions to the scattering amplitudes could be small because of decoupling, and the non-resonant search methods become relevant \cite{skalozub-dmytriiev}.

    It is convenient then to describe the interactions of  new particles with the low-energy effective Lagrangian (EL) of the SM fields, which consists of high-dimensional operators. Then contributions of these operators could be constrained by experiment. The low-energy effective Lagrangians of  new physics models  are different -- some types of operators are suppressed or enhanced in a particular model. Thus, it is necessary to obtain the experimental constraints for the effective Lagrangian of each model, to improve the experimental reach \cite{marzocca}. In the present paper, we derive the low-energy effective Lagrangian of the two-Higgs-doublet model (2HDM). A detailed review of this model could be found in \cite{ivanov, ginzburg-krawczyk, gunion-haber, ginzburg-osland, gunion-haber-kane}.

    Here we consider the 2HDM as one of the extensions of the SM, which introduces a wide variety of new phenomena. For instance, one of Sakharov's baryogenesis conditions could be fulfilled within the SM extended with one scalar doublet \cite{kozhushko-skalozub}. As it is known, the minimal SM does not possess  this feature \cite{rubakov}.

    The 2HDM predicts that there exist five "physical" scalar particles, while only one has been experimentally observed as the Higgs boson. We investigate the case when the SM Higgs boson is the lightest state of the 2HDM, and the other four scalar particles are heavy. We integrate over these heavy scalar bosons and obtain the low-energy EL of the 2HDM. We derive the analytical expressions for the parameters of the tree-level potential of the SM Higgs boson in this EL. Then we derive  new effective operators of dimensions $5$ and $6$. They are introduced by interactions with the heavy 2HDM bosons, and we find the analytical expressions for the couplings of these operators. All  the corrections we provide up to the order of $\Lambda^{-2}$, where $\Lambda$ is the mass scale of the heavy bosons. We point out decoupling phenomenon in the considered model.

    The scenario where some or all of the scalar bosons become heavy was considered in\,\cite{belusca, gunion-haber, ciafaloni}. CP-conserving potential of the 2HDM which is symmetric under the change of sign of one of the doublets was discussed in these papers. Expressions for the couplings of "physical" scalars to other fields were obtained in \cite{gunion-haber}. In that research, a scenario  where couplings between the non-minimal scalars and the SM particles are small for some values of the model parameters was discussed. Low-energy effective Lagrangian of the 2HDM was obtained in \cite{ciafaloni} for the case when all physical 2HDM particles are beyond the reach of the modern colliders. However, the discovery of the $125\,GeV$ Higgs boson makes this hypothesis questionable, so we do not proceed with it. Authors of \cite{belusca} have obtained the low-energy EL for the 2HDM where one of the scalars is light and the others are heavy. As it was shown there, such variant of the 2HDM does not fit  good enough to the LHC Higgs data, and some modifications of the model are required. In our paper, we choose the more general potential, discussed in \cite{ginzburg-krawczyk, ginzburg-osland}, which also allows for a small violation of CP-symmetry, and obtain the low-energy EL for such a model.

    This paper is organized as follows. In section \ref{sec:potential-of-the-two-higgs-doublet-model} we discuss the particle spectrum of the model and analyze properties of the particles. Section \ref{sec:low-energy-effective-lagrangian-of-the-two-higgs-doublet} contains the low-energy effective Lagrangian of the 2HDM. Also here we figure out parameters of the SM when heavy bosons decouple and couplings of the effective operators introduced by the 2HDM. Section \ref{sec:conclusions} summarises  our results. We provide analytical expressions for the mass matrices of the scalar particles and for the terms of Yukawa's interaction between the 2HDM scalars and the SM fermions in the Appendix.

    \section{Two-Higgs-doublet model potential}
    \label{sec:potential-of-the-two-higgs-doublet-model}
    We start with the Lagrangian of the 2HDM scalar fields $\mathcal{L}_s$:
    \begin{align}
    \label{scalar-lagrangian}
    \mathcal{L}_s = \sum\limits_{i=1,2}\left(D^{\mu}\phi_i\right)^{\dagger}D_{\mu}\phi_i - V(\phi_1;\phi_2),\quad iD_{\mu} = i\partial_{\mu} + \frac{1}{2}g\sigma_a W^a_{\mu} + \frac{1}{2}g^{\prime}B_{\mu},\quad a = \overline{1;3}.
    \end{align}
    Here $\phi_1$ and $\phi_2$ denote two scalar doublets. $V(\phi_1;\phi_2)$ is the potential of the scalar fields. There is also  Lagrangian $\mathcal{L}_Y$ of Yukawa's interaction between the scalar doublets and the SM fermions, which we discuss in  next section. In our investigation, we consider only the effective vertexes with the SM Higgs $h$ and/or fermions in the initial and final states. Contributions of the weak gauge bosons to these vertexes are of the next-to-leading order. So we neglect them and omit the gauge fields in the kinetic term in \eqref{scalar-lagrangian}.

    There are many possible types of interactions between particles which could be introduced by a general potential of the two-Higgs-doublet model. In our paper, we choose the specific potential
    \begin{align}
    \label{model-potential}
    V(\phi_1;\phi_2) &= m_{11}^2\phi_1^{\dagger}\phi_1 + m_{22}^2\phi_2^{\dagger}\phi_2 - (m_{12}^2\phi_1^{\dagger}\phi_2 + m_{12}^{2*}\phi_2^{\dagger}\phi_1) + \nonumber \\
    &+ \frac{1}{2}\lambda_1(\phi_1^{\dagger}\phi_1)^2 + \frac{1}{2}\lambda_2(\phi_2^{\dagger}\phi_2)^2 + \lambda_3(\phi_1^{\dagger}\phi_1)(\phi_2^{\dagger}\phi_2) + \lambda_4(\phi_1^{\dagger}\phi_2)(\phi_2^{\dagger}\phi_1) + \nonumber \\
    &+ \frac{1}{2}\left[\lambda_5(\phi_1^{\dagger}\phi_2)^2 + \lambda_5^*(\phi_2^{\dagger}\phi_1)^2\right],
    \end{align}
    \begin{equation}
    \label{initial-scalar-fields-parametrization}
    \phi_i = \begin{pmatrix}
    a_i^+ \\
    \phi_i^0
    \end{pmatrix},\quad \phi_i^0 = \frac{1}{\sqrt{2}}\left(v_i + b_i + ic_i\right).
    \end{equation}
    Here $a_i^+$, $b_i$ and $c_i$ are charged, neutral CP-even and neutral CP-odd components of the doublet $\phi_i$, respectively. Neutral components of the doublets have real vacuum expectation values (VEVs) $\frac{1}{\sqrt{2}}v_1$ and $\frac{1}{\sqrt{2}}v_2$, $v_1 > v_2$. All parameters in the potential \eqref{model-potential} are real, except $m_{12}^2$ and $\lambda_5$. Because of this, there are neutral scalars with unspecified CP-parity among the mass eigenstates of the model. The Yukawa interaction of these states with fermions violates CP-parity \cite{ginzburg-krawczyk}, and the magnitude of CP-violation is regulated by $\Im\,\lambda_5$.

    Vacuum state of the model minimizes the potential \eqref{model-potential}:
    \[\frac{\partial V}{\partial \phi_1}\Bigr|_{vac} = 0,\quad \frac{\partial V}{\partial \phi_2}\Bigr|_{vac} = 0.\]
    From these equalities we find the relations between some of the model parameters:
    \begin{align*}
    m_{11}^2 &= \frac{v_2}{v_1}\Re\,m_{12}^2 - \frac{1}{2}\left[\lambda_1 v_1^2 + v_2^2\left(\lambda_3 + \lambda_4 + \Re\,\lambda_5\right)\right], \\
    m_{22}^2 &= \frac{v_1}{v_2}\Re\,m_{12}^2 - \frac{1}{2}\left[\lambda_2 v_2^2 + v_1^2\left(\lambda_3 + \lambda_4 + \Re\,\lambda_5\right)\right], \\
    \Im\,m_{12}^2 &= \frac{1}{2}v_1 v_2 \Im\,\lambda_5.
    \end{align*}
    We investigate the scenario when one of the mass eigenstates has the same mass as the SM Higgs boson, and four other are very heavy and decouple at energies of order $O(v)$, where $v$ is the SM Higgs VEV. This scenario could be realized if we put $\Re\,m_{12}^2$ to be very big \cite{belusca, gunion-haber, gunion-haber-kane}. In what follows, we consider $\Re\,m_{12}^2$, $v_1$, $v_2$ and scalar self-couplings $\lambda_i$, $i=\overline{1;5}$ as free parameters of the model. For simplicity we also assume that $\Im\,\lambda_5$ is small.

    The mass matrices of  scalar fields are given by the coefficients in the quadratic terms of the Taylor series expansion of \eqref{model-potential} near its minimum,
    \begin{equation*}
    V(\phi_1;\phi_2) = V(\phi_1;\phi_2)\Bigr|_{vac} + \begin{pmatrix}
    a_1^+ \\ a_2^+
    \end{pmatrix}^T M_a^2\begin{pmatrix}
    a_1^- \\ a_2^-
    \end{pmatrix} + \frac{1}{2}\begin{pmatrix}
    b_1 \\ b_2 \\ c_1 \\ c_2
    \end{pmatrix}^T M_{bc}^2\begin{pmatrix}
    b_1 \\ b_2 \\ c_1 \\ c_2
    \end{pmatrix} + O(\phi^3).
    \end{equation*}
    In this equation, $M_a^2$ and $M_{bc}^2$ are the mass matrices of the particles $a_i^+$, $b_i$ and $c_i$, respectively \cite{ginzburg-krawczyk, ginzburg-osland}. The expressions for them are given in the Appendix. Eigenstates of the matrix $M_a^2$ are the charged Goldstone boson $G^+$ and the massive particle $H^+$:
    \begin{align}
    \label{charged-mass-eigenstates-def}
    H^+ &= -a_1^+ \sin{\beta} + a_2^+ \cos{\beta},\quad G^+ = a_1^+ \cos{\beta} + a_2^+ \sin{\beta}, \nonumber \\
    \tan{\beta} &= \frac{v_2}{v_1}.
    \end{align}

    One of the eigenvalues of $M_{bc}^2$ is zero, so that there are three massive scalars $h_1$, $h_2$, $h_3$ and one Goldstone boson $G_0$:
    \begin{align*}
    \begin{pmatrix}
    h_1 \\ h_2 \\ h_3 \\ G_0
    \end{pmatrix} &= R \begin{pmatrix}
    b_1 \\ b_2 \\ c_1 \\ c_2
    \end{pmatrix},\quad R = \begin{pmatrix}
    1 & 0 \\
    0 & R_{\beta}
    \end{pmatrix},\quad R_{\beta} = \begin{pmatrix}
    -s_{\beta} & c_{\beta} \\
    c_{\beta} & s_{\beta}
    \end{pmatrix}, \nonumber \\
    s_{\beta} &= \sin{\beta},\quad c_{\beta} = \cos{\beta},\quad t_{\beta} = \tan{\beta}.
    \end{align*}
    The mass matrix of the massive neutral scalars $h_1$, $h_2$ and $h_3$ is
    \begin{align*}
    M_h^2 &= \begin{pmatrix}
    \lambda_1 v^2 c_{\beta}^2 + t_{\beta}\Re\,m_{12}^2 & v^2 s_{\beta}c_{\beta}\lambda_{345} - \Re\,m_{12}^2 & -\frac{1}{2}\Im\,\lambda_5 v^2 s_{\beta} \\
    v^2 s_{\beta}c_{\beta}\lambda_{345} - \Re\,m_{12}^2 & \lambda_2 v^2 s_{\beta}^2 + \frac{1}{t_{\beta}}\Re\,m_{12}^2 & -\frac{1}{2}\Im\,\lambda_5 v^2c_{\beta} \\
    -\frac{1}{2}\Im\,\lambda_5 v^2 s_{\beta} & -\frac{1}{2}\Im\,\lambda_5 v^2c_{\beta} & \frac{1}{s_{\beta}c_{\beta}}\Re\,m_{12}^2 - \Re\,\lambda_5 v^2
    \end{pmatrix}, \\
    v^2 &= v_1^2 + v_2^2,\quad \lambda_{345} = \lambda_3 + \lambda_4 + \Re\,\lambda_5.
    \end{align*}
    We diagonalize this matrix via  the $3$ rotations of the basis $\{h_1;h_2;h_3\}$ \cite{ginzburg-krawczyk}. The corresponding matrices  are
    \begin{align}
    \label{first-basis-rotation-def}
    R_1 &= \begin{pmatrix}
    c_{\alpha_1} & s_{\alpha_1} & 0 \\
    -s_{\alpha_1} & c_{\alpha_1} & 0 \\
    0 & 0 & 1
    \end{pmatrix},\quad R_2 = \begin{pmatrix}
    c_{\alpha_2} & 0 & s_{\alpha_2} \\
    0 & 1 & 0 \\
    -s_{\alpha_2} & 0 & c_{\alpha_2}
    \end{pmatrix},\quad R_3 = \begin{pmatrix}
    1 & 0 & 0 \\
    0 & c_{\alpha_3} & s_{\alpha_3} \\
    0 & -s_{\alpha_3} & c_{\alpha_3}
    \end{pmatrix}, \nonumber \\
    \alpha &= \alpha_1 - \frac{\pi}{2}.
    \end{align}
    Here we follow the notation of \cite{ginzburg-krawczyk} and use the angle $\alpha$ instead of $\alpha_1$. Non-diagonal element $M_{h 12}^2$ of $M_h^2$ vanishes after the rotation $R_1$:
    \begin{align*}
    M_h^{2\prime} = R_1 M_h^2 R_1^T = \begin{pmatrix}
    S + \Delta & 0 & -\frac{1}{2}\Im\,\lambda_5 v^2 c_{\alpha + \beta} \\
    0 & S - \Delta & \frac{1}{2}\Im\,\lambda_5 v^2 s_{\alpha + \beta} \\
    -\frac{1}{2}\Im\,\lambda_5 v^2 c_{\alpha + \beta} & \frac{1}{2}\Im\,\lambda_5 v^2 s_{\alpha + \beta} & \frac{1}{s_{\beta}c_{\beta}}\Re\,m_{12}^2 - \Re\,\lambda_5 v^2
    \end{pmatrix},
    \end{align*}
    where $S$, $\Delta$ and $\alpha$ are defined as
    \begin{align}
    \label{s-delta-def}
    S &= \frac{1}{2}\left[\frac{1}{s_{\beta}c_{\beta}}\Re\,m_{12}^2 + v^2(\lambda_1 c_{\beta}^2 + \lambda_2 s_{\beta}^2)\right], \nonumber \\
    \Delta &= \frac{1}{2\cos{2\alpha}}\left[\frac{2}{t_{2\beta}}\Re\,m_{12}^2 - v^2(\lambda_1 c_{\beta}^2 - \lambda_2 s_{\beta}^2)\right], 	
    \end{align}
    \begin{equation}
    \label{alpha-mixing-angle-def}
    \tan{2\alpha} = t_{2\beta}\frac{1 - \frac{1}{2}\varepsilon \lambda_{345}s_{2\beta}}{1 - \frac{1}{2}\varepsilon t_{2\beta}(\lambda_1 c_{\beta}^2 - \lambda_2 s_{\beta}^2)},\quad \varepsilon = \frac{v^2}{\Re\,m_{12}^2}.
    \end{equation}
    The angle $\alpha$ is such that $\cos{2\alpha} > 0$ by definition. We diagonalize $M_h^{2\prime}$ with rotations $R_2$ and $R_3$. Since $|\Im\,\lambda_5|\ll 1$ and $\alpha_2\sim \Im\,\lambda_5$, $\alpha_3\sim \Im\,\lambda_5$, the corresponding rotation angles $\alpha_2$ and $\alpha_3$ are small, too. So the following approximations for $R_2$ and $R_3$ are valid,
    \[R_2\approx \begin{pmatrix}
    1 & 0 & \alpha_2 \\
    0 & 1 & 0 \\
    -\alpha_2 & 0 & 1
    \end{pmatrix},\quad R_3\approx \begin{pmatrix}
    1 & 0 & 0 \\
    0 & 1 & \alpha_3 \\
    0 & -\alpha_3 & 1
    \end{pmatrix}.\]
    We neglect all the terms of the second and higher orders in $\Im\,\lambda_5$, and obtain the following approximations for $\alpha_2$ and $\alpha_3$:
    \begin{align}
    \label{final-diagonalization-angles-def}
    \alpha_2 &\approx \frac{\Im\,\lambda_5 v^2\cos{(\alpha + \beta)}}{2M_{h33}^2 - 2(S + \Delta)}, \nonumber \\
    \alpha_3 &\approx -\frac{\Im\,\lambda_5 v^2 \sin{(\alpha + \beta)}}{2M_{h33}^2 - 2(S - \Delta)}.
    \end{align}
    Here $M_{h33}^2$ denotes the third diagonal element of the mass matrix $M_h^2$. Finally, we obtain the neutral mass eigenstates $H$, $h$ and $A_0$
    \begin{align}
    \label{neutral-mass-eigenstates-def}
    \begin{pmatrix}
    H \\ h \\ A_0
    \end{pmatrix} &= R_3 R_2 R_1\begin{pmatrix}
    h_1 \\ h_2 \\ h_3
    \end{pmatrix} = \begin{pmatrix}
    -b_1 s_{\alpha} + b_2 c_{\alpha} + \alpha_2\left(c_2 \cos{\beta} - c_1 \sin{\beta}\right) \\
    -b_1 c_{\alpha} - b_2 s_{\alpha} + \alpha_3\left(c_2 \cos{\beta} - c_1 \sin{\beta}\right) \\
    b_1\left(\alpha_3 c_{\alpha} + \alpha_2 s_{\alpha}\right) + b_2\left(- \alpha_2 c_{\alpha} + \alpha_3 s_{\alpha}\right) + c_2 \cos{\beta} - c_1 \sin{\beta}
    \end{pmatrix}, \nonumber \\
    G_0 &= c_1\cos{\beta} + c_2\sin{\beta}.
    \end{align}
    As one can see from these expressions, the neutral fields $h$, $H$ and $A_0$ do not have definite CP-parities, because they are linear combinations of the CP-even fields $b_1$, $b_2$ and CP-odd fields $c_1$ and $c_2$. This mixing is proportional to $\Im\,\lambda_5$. So, when $\Im\,\lambda_5 = 0$, $h$ and $H$ become CP-even, and $A_0$ becomes CP-odd. Simultaneously, CP-parity of the Goldstone boson $G_0$ does not depend on $\Im\,\lambda_5$, and this particle always remains CP-odd.

    The masses of the particles are given in the table\,\ref{table:mass-eigenstates-masses}.
    \begin{table}[h]
    	\caption{Masses of the 2HDM bosons}
    	\begin{center}
    		\begin{tabular}{|c|c|}
    			\hline
    			$H^{\pm}$ & $m_{H^+}^2 = \cfrac{1}{s_{\beta}c_{\beta}}\Re\,m_{12}^2 - \cfrac{1}{2}v^2(\lambda_4 + \Re\,\lambda_5)$ \\
    			\hline
    			$A_0$ & $m_A^2 = \cfrac{1}{s_{\beta}c_{\beta}}\Re\,m_{12}^2 - \Re\,\lambda_5 v^2$ \\
    			\hline
    			$H$ & $m_H^2 = S + \Delta$ \\
    			\hline
    			$h$ & $m_h^2 = S - \Delta$ \\
    			\hline
    		\end{tabular}
    	\end{center}
    	\label{table:mass-eigenstates-masses}
    \end{table}

    In the 2HDM, the masses of the weak gauge bosons  are introduced by interaction with the scalar doublets, and they are proportional to $v$. Hence, $v$ equals  to the VEV of the Higgs field in the minimal SM -- $v\approx 250\,GeV$. In the limit when $\Re\,m_{12}^2 \gg v^2$ and all scalar self-couplings are $\sim O(1)$, the  particles $H^{\pm}$, $H$ and $A_0$ become heavy and nearly degenerate in masses, as it is shown in the table\,\ref{table:mass-eigenstates-big-re-m12}.

    \begin{table}[h]
    	\caption{Masses of the Higgs bosons in the limit $\Re\,m_{12}^2 \gg v^2$}
    	\begin{center}
    		\begin{tabular}{|c|c|}
    			\hline
    			$H^{\pm}$ & $m_{H^+}^2 \approx \cfrac{1}{s_{\beta}c_{\beta}}\Re\,m_{12}^2$ \\
    			\hline
    			$A_0$ & $m_A^2 \approx \cfrac{1}{s_{\beta}c_{\beta}}\Re\,m_{12}^2$ \\
    			\hline
    			$H$ & $m_H^2 \approx \cfrac{1}{s_{\beta}c_{\beta}}\Re\,m_{12}^2$ \\
    			\hline
    			$h$ & $m_h^2 \approx \cfrac{1}{2}\lambda_1 v^2 c_{\beta}^2\left(1 + \cfrac{1}{\cos{2\beta}}\right) + \cfrac{1}{2}\lambda_2 v^2 s_{\beta}^2\left(1 - \cfrac{1}{\cos{2\beta}}\right)$ \\
    			\hline
    		\end{tabular}
    	\end{center}
    	\label{table:mass-eigenstates-big-re-m12}
    \end{table}

    This limit also implies that $\tan{2\alpha}\rightarrow \tan{2\beta}$. Mass of the scalar boson $h$ is then $O(v)$, so this quantity could be close to that of the SM Higgs boson.

    Hereafter we use the following parametrization for the scalar doublets:
    \begin{align}
    \label{unitary-gauge-def}
    \phi_1 &= U\begin{pmatrix}
    -H^+ s_{\beta} \\
    \frac{1}{\sqrt{2}}\left[v_1 + (A_0\alpha_3 - h)c_{\alpha} + (A_0\alpha_2 - H)s_{\alpha} - i(A_0 + H\alpha_2 + h\alpha_3)s_{\beta}\right]
    \end{pmatrix}, \nonumber \\
    \phi_2 &= U\begin{pmatrix}
    H^+ c_{\beta} \\
    \frac{1}{\sqrt{2}}\left[v_2 + (A_0\alpha_3 - h)s_{\alpha} - (A_0\alpha_2 - H)c_{\alpha} + i(A_0 + H\alpha_2 + h\alpha_3)c_{\beta}\right]
    \end{pmatrix}, \nonumber \\
    U &= \exp\left[-i\frac{(\overline{G}\overline{\sigma})}{v}\right],\quad (\overline{G}\overline{\sigma}) = \sigma^1 G_1 + \sigma^2 G_2 + \sigma^3 G_3, \nonumber \\
    G^{\pm} &= \frac{1}{\sqrt{2}}(G_2 \mp iG_1),\quad G_0 = G_3.
    \end{align}
    Here $\sigma_a$, $a=\overline{1;3}$ denote Pauli's matrices. The original parametrization \eqref{initial-scalar-fields-parametrization} could be obtained from \eqref{unitary-gauge-def} if one neglects the terms which are quadratic in fields \cite{ciafaloni}. In the unitary gauge \eqref{unitary-gauge-def}, the inessential Goldstone degrees of freedom do not enter the potential \eqref{model-potential}, and $V(\phi_1;\phi_2)$ is represented in terms of the "physical" scalar fields, only.

    \section{Low-energy effective Lagrangian of the 2HDM}
    \label{sec:low-energy-effective-lagrangian-of-the-two-higgs-doublet}

    We assume in our treatment that the SM Higgs is the lightest scalar boson of the Standard model with two scalar doublets, and it is described with the $h$ field. Then the high-energy dynamics of the 2HDM and fermions is described by the following Lagrangian:
    \begin{equation}
    \label{high-energy-lagrangian}
    \mathcal{L} = \mathcal{L}_s + \mathcal{L}_Y.
    \end{equation}
    The second term of \eqref{high-energy-lagrangian}, $\mathcal{L}_Y$, is the Lagrangian of Yukawa's interaction:
    \begin{align}
    \label{yukawa-lagrangian}
    \mathcal{L}_Y = -\sum\limits_{f;f^{\prime}}\sum\limits_{i=1,2}&\left\{y_{ff^{\prime}}^{i(1)(q)}(\overline{Q}_L^{(f)}\phi_i)d_R^{(f^{\prime})} + y_{ff^{\prime}}^{i(2)(q)}(\overline{Q}_L^{(f)}\phi_i^c)u_R^{(f^{\prime})} +\right. \nonumber \\
    &\left.+ y_{ff^{\prime}}^{i(1)(l)}(\overline{L}_L^{(f)}\phi_i)e_R^{(f^{\prime})} + y_{ff^{\prime}}^{i(2)(l)}(\overline{L}_L^{(f)}\phi_i^c)\nu_R^{(f^{\prime})} + h.\,c.\right\}, \nonumber \\
    f;f^{\prime} &= \overline{1;3},\quad \phi_i^c = -i\sigma_2\phi_i^*,\quad Q_L^{(f)} = \begin{pmatrix}
    u_L^{(f)} \\
    d_L^{(f)}
    \end{pmatrix},\quad L_L^{(f)} = \begin{pmatrix}
    \nu_L^{(f)} \\
    e_L^{(f)}
    \end{pmatrix}.
    \end{align}
    In this expression $y_{ff^{\prime}}^{(1)(q)}$, $y_{ff^{\prime}}^{(2)(q)}$, $y_{ff^{\prime}}^{(1)(l)}$ and $y_{ff^{\prime}}^{(2)(l)}$ are the Yukawa couplings. Superscripts $(q)$ and $(l)$ denote couplings which describe interactions with quarks and leptons, respectively. $\phi_i^c$ is the doublet which is charge-conjugated to $\phi_i$, $Q_L^{(f)}$, and $L_L^{(f)}$ are the doublets of left-handed quarks and leptons of a generation $f$, respectively. For instance, $u_L^{(1)}$ is a left-handed $u$-quark, $d_R^{(2)}$ is a right-handed $s$-quark etc. Similarly, $\nu_L^{(1)}$ is a left-handed electron neutrino, and $e_L^{(3)}$ is a left-handed tau-lepton. Fermion doublets in \eqref{yukawa-lagrangian} are parametrized in such a gauge that Goldstone's bosons do not enter $\mathcal{L}_Y$. Besides, all fermionic fields in \eqref{yukawa-lagrangian} are symmetry eigenstates.

    As it is known from the experimental data, there are no tree-level flavour-changing interactions between charged leptons or quarks of the same charge, in the considered range of energies. This fact could be taken into account with the specific choice of pattern or values of the Yukawa couplings. However, the main results of our investigation do not depend on such constraints. So we use the general expression for the Yukawa Lagrangian \eqref{yukawa-lagrangian}.

    In terms of the mass eigenstates of the 2HDM, the Yukawa Lagrangian \eqref{yukawa-lagrangian} reads
    \begin{align}
    \label{yukawa-lagrangian-mass-eigenstates}
    \mathcal{L}_Y = -J^+ H^- - J^- H^+ - J_H H - J_A A_0 - J_h h.
    \end{align}
    In this equation, $J^+$, $J^-$, $J_H$, $J_A$ and $J_h$ denote the contributions of the SM fermionic fields,
    \begin{equation}
    \label{yukawa-interaction-terms}
    J^{\pm} = J^{\pm (q)} + J^{\pm (l)},\quad J_H = J_H^{(q)} + J_H^{(l)},\quad J_A = J_A^{(q)} + J_A^{(l)},\quad J_h = J_h^{(q)} + J_h^{(l)}.
    \end{equation}
    Here operators $J^{\pm(q)}$, $J_H^{(q)}$, $J_A^{(q)}$ and $J_h^{(q)}$ contain quark fields, while $J^{\pm(l)}$, $J_H^{(l)}$, $J_A^{(l)}$ and $J_h^{(l)}$ consist of leptonic ones. These terms are adduced in  Appendix.

    Lagrangian \eqref{scalar-lagrangian} in terms of the mass eigenstates is 
    \begin{align}
    \label{explicit-scalar-lagrangian}
    \mathcal{L}_s &= \frac{1}{2}\sum\limits_{a = 1}^3(\partial_{\mu}G_a)^2 + \partial^{\mu}H^+\partial_{\mu}H^- + \frac{1}{2}(\partial_{\mu}A_0)^2 + \nonumber \\
    &+ \frac{1}{2}(\partial_{\mu}H)^2 + \frac{1}{2}(\partial_{\mu}h)^2 - V(H^{\pm};A_0;H;h).
    \end{align}
    Hereafter we enumerate fields $H^+$, $H^-$, $H$ and $A_0$ with one latin index:
    \[\{H^+;H^-;H;A_0\} = \{H^a\},\quad a=\overline{1;4}.\]

    Now we derive the effective action $\Gamma_{eff}$ of the light particles of the theory. $\Gamma_{eff}$ describes the interactions of the light particles in the processes where the non-minimal Higgs bosons $H^{\pm}$, $H$ and $A_0$ do not appear in the initial or final states. Instead, they participate in the interactions  as virtual states only, and contribute the low-energy dynamics via the effective operators of the SM fields. We integrate over the non-minimal scalar bosons and derive $\Gamma_{eff}$:
    \begin{equation}
    \label{effective-action-def}
    e^{i\Gamma_{eff}} = \int\mathcal{D}H\mathcal{D}A_0\mathcal{D}H^+\mathcal{D}H^-\exp{\left(i\int d^4x \mathcal{L}\right)}
    \end{equation}
    Since $\mathcal{L}$ contains terms which are cubic and quartic in the scalar fields, we calculate $\Gamma_{eff}$ in the Gaussian approximation. That is, we expand action of the scalar fields near some classical field configuration $H_{class}^a$,
    \begin{align}
    \label{total-action-gaussian-approximation}
    S[h;H^a] &= \int d^4x \mathcal{L} = S[h;H_{class}^a] + \int d^4x\frac{\delta S}{\delta H^a(x)}\Bigr|_{H^a = H_{class}^a}\Delta H^a(x) + \nonumber \\
    &+ \frac{1}{2}\int d^4x_1 d^4x_2 \frac{\delta^2 S}{\delta H^a(x_1)\delta H^b(x_2)}\Bigr|_{H^a = H_{class}^a}\Delta H^a(x_1)\Delta H^b(x_2) + O((\Delta H^a)^3), \nonumber \\
    \Delta H^a(x) &= H^a(x) - H_{class}^a(x).
    \end{align}
    The  fields $H_{class}^a$ are such that $S$ has a minimum at $H_{class}^a$, and we find this configuration as the solution to the classical motion equations
    \begin{equation}
    \label{classical-motion-equations}
    \frac{\delta S}{\delta H^a(x)}\Bigr|_{H^a = H_{class}^a} = 0 \Rightarrow \partial^2 H_{class}^a + \frac{\partial V}{\partial H^a}\Bigr|_{H^a = H_{class}^a} - \frac{\partial \mathcal{L}_Y}{\partial H^a}\Bigr|_{H^a = H_{class}^a} = 0.
    \end{equation}
    Simultaneously, we neglect all of the terms which contain $\Delta H^a(x)$ in powers which are bigger than two in the expansion \eqref{total-action-gaussian-approximation}. In this way the effective action $\Gamma_{eff}$ accounts only for the contributions of small quantum fluctuations over the classical background $H_{class}^a$.

    Classical motion equations \eqref{classical-motion-equations} are non-linear, and we solve them approximately, similarly to \cite{belusca}. In the zeroth order in the scalar self-couplings and for energies $|p^2|\ll \Re\,m_{12}^2$, $|p^2| = O(v^2)$, the solutions are
    \begin{align}
    \label{classical-motion-equations-solutions}
    H_{class}^{\pm} &\approx -\frac{1}{m_{H^+}^2}J^{\pm},\quad A_{0 class} \approx -\frac{1}{m_A^2}J_A,\quad H_{class} \approx -\frac{1}{m_H^2}J_H, \nonumber \\
    m_{H^+}^2 &\approx m_A^2 \approx m_H^2 \approx \frac{1}{s_{\beta}c_{\beta}}\Re{m_{12}^2}\Rightarrow \nonumber \\
    &\Rightarrow H_{class}^{\pm} \approx -s_{\beta}c_{\beta}\frac{\varepsilon}{v^2}J^{\pm},\quad A_{0 class} \approx -s_{\beta}c_{\beta}\frac{\varepsilon}{v^2}J_A,\quad H_{class} \approx -s_{\beta}c_{\beta}\frac{\varepsilon}{v^2}J_H.
    \end{align}
    Here we neglected the kinetic terms in the equations \eqref{classical-motion-equations} within the low-energy approximation
    \begin{align*}
    |p^2|\ll |\Re\,m_{12}^2|\Rightarrow |\partial^2 H^{\pm}| \ll \Re\,m_{12}^2 |H^{\pm}|,\quad |\partial^2 H| \ll \Re\,m_{12}^2 |H|,\quad |\partial^2 A_0| \ll \Re\,m_{12}^2 |A_0|.
    \end{align*}
    We insert the solutions \eqref{classical-motion-equations-solutions} into the Lagrangians $\mathcal{L}_s$ and $\mathcal{L}_Y$, and get
    \begin{align}
    \label{zero-loop-effective-scalar-lagrangian}
    \mathcal{L}_s[h;H_{class}^a] &= \frac{1}{2}\sum\limits_{a = 1}^3 (\partial_{\mu} G_a)^2 + \frac{1}{2}(\partial_{\mu}h)^2 - \frac{1}{2}m_h^2 h^2 - \lambda^{(3)}h^3 - \lambda^{(4)}h^4 - \nonumber \\
    &- \frac{\varepsilon}{v}(C_2 J_A + C_3 J_H)h^2 - \frac{\varepsilon}{v^2}(C_4 J_A + C_5 J_H) h^3,
    \end{align}
    \vspace{-5mm}
    \begin{align}
    \label{zero-loop-effective-yukawa-lagrangian}
    \mathcal{L}_Y[h;H_{class}^a] &= -J_h h + s_{\beta}c_{\beta}\frac{\varepsilon}{v^2}\left(2J^+ J^- + J_H^2 + J_A^2\right).
    \end{align}
    Here we have taken into consideration only the operators which have canonical dimensions less than seven and neglected the others, which are suppressed by such factors as $(\Re\,m_{12}^2)^{-d}$, $d\geq 2$. $\lambda^{(3)}$, $\lambda^{(4)}$ and $C_i$, $i=\overline{2;5}$ are constants. Their values are 
    \begin{align}
    \label{sm-higgs-potential-coefficients}
    \lambda^{(3)} &= -\frac{1}{2}v\left(\lambda_1 c_{\alpha}^3 c_{\beta} + \lambda_2 s_{\alpha}^3 s_{\beta} + \frac{1}{2}\lambda_{345}s_{2\alpha}s_{\alpha + \beta}\right), \nonumber \\
    \lambda^{(4)} &= \frac{1}{8}\left(\lambda_1 c_{\alpha}^4 + \frac{1}{2}\lambda_{345}s_{2\alpha}^2 + \lambda_2 s_{\alpha}^4\right),
    \end{align}
    \vspace{-5mm}
    \begin{align}
    \label{fermion-higgs-effective-couplings-h-square}
    C_2 &= \frac{1}{4}\left[\Im\,\lambda_5 s_{2\beta}\left(\frac{1}{2}s_{2\beta} + s_{2\alpha}\right) +\right. \nonumber \\
    &\left.+ \alpha_2\left(-\frac{3}{2}s_{2\alpha}s_{2\beta}(\lambda_1 c_{\alpha}c_{\beta} - \lambda_2 s_{\alpha}s_{\beta}) + \lambda_{345}s_{2\alpha}s_{2\beta}c_{\alpha + \beta} + \lambda_{345}s_{2\beta}(s_{\beta}c_{\alpha}^3 - c_{\beta}s_{\alpha}^3)\right) +\right. \nonumber \\
    &+ 2\alpha_3\Bigl(\lambda_1 s_{\beta}c_{\beta}^2 c_{\alpha}(2s_{\beta}^2 - 3c_{\alpha}^2) + \lambda_2 c_{\beta}s_{\alpha}s_{\beta}^2(2c_{\beta}^2 - 3s_{\alpha}^2) - \nonumber \\
    &\left.\left.- 2\Re\,\lambda_5 s_{2\beta}c_{\alpha - \beta} + \lambda_{345}\left(s_{2\beta}(c_{\alpha} c_{\beta}^3 + s_{\alpha} s_{\beta}^3) - \frac{3}{4}s_{2\alpha}s_{2\beta}s_{\alpha + \beta}\right)\right)\right], \nonumber \\
    C_3 &= \frac{1}{4}\left[\frac{3}{2}s_{2\alpha}s_{2\beta}\left(\lambda_1 c_{\alpha}c_{\beta} - \lambda_2 s_{\alpha} s_{\beta}\right) - \lambda_{345}\left(s_{2\alpha}s_{2\beta}c_{\alpha + \beta} - s_{2\beta}(c_{\beta}s_{\alpha}^3 - s_{\beta}c_{\alpha}^3)\right)\right].
    \end{align}
    \vspace{-5mm}
    \begin{align}
    \label{fermion-higgs-effective-couplings-h-cube}
    C_4 =& -\frac{1}{8}\Im\,\lambda_5 s_{2\alpha} s_{2\beta} c_{\alpha - \beta} + \frac{\alpha_2}{8}s_{2\alpha}s_{2\beta}\left(\lambda_1 c_{\alpha}^2 - \lambda_2 s_{\alpha}^2 - \lambda_{345}c_{2\alpha}\right) + \nonumber \\
    +& \frac{\alpha_3}{2}\left[\frac{1}{2}\lambda_1 s_{2\beta}c_{\alpha}^2(c_{\alpha}^2 - s_{\beta}^2) + \frac{1}{2}\lambda_2 s_{2\beta} s_{\alpha}^2 (s_{\alpha}^2 - c_{\beta}^2) +\right. \nonumber \\
    &\left.+ \Re\,\lambda_5 s_{2\beta}c_{\alpha - \beta}^2 - \frac{1}{2}\lambda_{345} s_{2\beta}(c_{\alpha}^2 (c_{\beta}^2 - s_{\alpha}^2) + s_{\alpha}^2 (s_{\beta}^2 - c_{\alpha}^2))\right], \nonumber \\
    C_5 =& -\frac{1}{8}s_{2\alpha}s_{2\beta}\left[(\lambda_1 c_{\alpha}^2 - \lambda_2 s_{\alpha}^2) - \lambda_{345}c_{2\alpha}\right].
    \end{align}

    Now we turn to the contribution of the quadratic terms in  Gaussian's approximation \eqref{total-action-gaussian-approximation}. These terms could be represented in the following matrix form:
    \begin{align}
    \label{action-second-functional-derivative}
    &\frac{\delta^2 S}{\delta H^a(x_1)\delta H^b(x_2)}\Bigr|_{H^a = H_{class}^a}\Delta H^a(x_1)\Delta H^b(x_2) = \begin{pmatrix}
    \Delta H^+ \\ \Delta H^- \\ \Delta H \\ \Delta A_0
    \end{pmatrix}^T M_S \begin{pmatrix}
    \Delta H^+ \\ \Delta H^- \\ \Delta H \\ \Delta A_0
    \end{pmatrix}, \nonumber \\
    &M_S = M_S^{(0)} + \delta M_S,\quad M_S^{(0)} = -\begin{pmatrix}
    0 & \partial^2 + m_{H^+}^2 & 0 & 0 \\
    \partial^2 + m_{H^+}^2 & 0 & 0 & 0 \\
    0 & 0 & \partial^2 + m_H^2 & 0 \\
    0 & 0 & 0 & \partial^2 + m_A^2
    \end{pmatrix}, \nonumber \\
    &\delta M_S = -\begin{pmatrix}
    0 & \delta^{+-} & \delta_H^+ & \delta_A^+ \\
    \delta^{+-} & 0 & \delta_H^- & \delta_A^- \\
    \delta_H^+ & \delta_H^- & \delta_{HH} & \delta_{HA} \\
    \delta_A^+ & \delta_A^- & \delta_{HA} & \delta_{AA}
    \end{pmatrix}.
    \end{align}
    As we can see here, the matrix $M_S^{(0)}$ contains inverse propagators of  free fields $H^{\pm}$, $H$ and $A_0$. The functional integral of exponent of the quadratic terms is
    \begin{align}
    \label{quadratic-part-functional-integral}
    &\int\mathcal{D}H\mathcal{D}A_0\mathcal{D}H^+\mathcal{D}H^-\exp\left[\frac{i}{2}\int d^4x_1 d^4x_2\frac{\delta^2 S}{\delta H^a(x_1)\delta H^b(x_2)}\Bigr|_{H^a = H_{class}^a}\Delta H^a(x_1)\Delta H^b(x_2)\right] = \nonumber \\
    &= \left(\det{M_S}\right)^{-\frac{1}{2}} = \exp\left[-\frac{1}{2}\Tr\ln{M_S^{(0)}} - \frac{1}{2}\Tr\ln\left(1 + M_S^{(0)-1}\delta M_S\right)\right], \nonumber \\
    &M_S^{(0)} = \begin{pmatrix}
    0 & G^{\pm -1} & 0 & 0 \\
    G^{\pm -1} & 0 & 0 & 0 \\
    0 & 0 & G_H^{-1} & 0 \\
    0 & 0 & 0 & G_A^{-1}
    \end{pmatrix}.
    \end{align}
    In this equation, trace is computed over both spatial and discrete indices of the matrix $M_S$. The term $\Tr\ln{M_S^{(0)}}$ does not contain any fields and is constant, so we omit it. The matrix $\delta M_S$ consists of the terms which come from the self-interaction part of the potential \eqref{model-potential}. The Taylor series  of the logarithm in \eqref{quadratic-part-functional-integral} in powers of $M_S^{(0)-1}\delta M_S$ is equivalent to a perturbative series expansion. In the first order in the scalar self-couplings, the last term in the square brackets in \eqref{quadratic-part-functional-integral} equals to 
    \begin{align}
    \label{one-loop-effective-scalar-lagrangian}
    &\exp\left[-\frac{1}{2}\Tr\ln\left(1 + M_S^{(0)-1}\delta M_S\right)\right]\approx \exp\left[-\frac{1}{2}\Tr\left(M_S^{(0)-1}\delta M_S\right)\right] = \nonumber \\
    &= \exp\left[\frac{1}{2}\int d^4x \left(2G^{\pm}(x;x)\delta^{+-}(x) + G_H(x;x)\delta_{HH}(x) + G_A(x;x)\delta_{AA}(x)\right)\right].
    \end{align}
    Here $\delta^{\pm}(x)$, $\delta_{HH}(x)$ and $\delta_{AA}(x)$ contain only the terms which are proportional to $J_H$, $J_A$, $J_H h$ and $J_A h$. $G^{\pm}(x;x)$, $G_H(x;x)$, and $G_A(x;x)$ are the constants which describe contributions of   heavy scalar loops. We include these terms into the renormalization of fermionic masses and the corresponding Yukawa couplings. So their contributions are not observable.

    Finally, the effective Lagrangian of the 2HDM is obtained,
    \begin{align}
    \label{effective-lagrangian}
    \mathcal{L}_{eff} &= \frac{1}{2}\sum\limits_{a = 1}^3 (\partial_{\mu} G_a)^2 + \frac{1}{2}(\partial_{\mu}h)^2 - \frac{1}{2}m_h^2 h^2 - \lambda^{(3)}h^3 - \lambda^{(4)}h^4 - J_h h - \nonumber \\
    &- \frac{\varepsilon}{v}(C_2 J_A + C_3 J_H)h^2 - \frac{\varepsilon}{v^2}(C_4 J_A + C_5 J_H) h^3 + s_{\beta}c_{\beta}\frac{\varepsilon}{v^2}\left(2J^+ J^- + J_H^2 + J_A^2\right).
    \end{align}
    The first line of this expression describes dynamics of the neutral scalar $h$ and its interaction with fermions. The second line contains the effective contact interactions which are introduced by the extra heavy scalars of the 2HDM at energies much below $\Re\,m_{12}^2$. These effective interactions are suppressed by the term $\left(\Re\,m_{12}^2\right)^{-1}$.

    Let us consider the decoupling limit of $\mathcal{L}_{eff}$, when $\varepsilon\rightarrow 0$. In this limit the contact interactions vanish, and Yukawa sector of the model is the same as that in the SM. For instance, in this limit fermionic masses could be explained by the spontaneous symmetry breaking in the scalar sector with one doublet\footnote{For the expressions of $J_h$ and the fermionic mass terms see \eqref{quark-scalar-current-light-h} and \eqref{fermionic-mass-terms} in the Appendix}.

    At the same time, when $\varepsilon \rightarrow 0$ the $\mathcal{L}_s$ does not coincide with the Lagrangian of one-Higgs-doublet model. In this limit the relation between $m_h^2$ and self-couplings $\lambda^{(3)}$ and $\lambda^{(4)}$ is different from that in the SM. Indeed, when $\varepsilon \rightarrow 0$ we have the following relations for the mixing angles, using definitions \eqref{alpha-mixing-angle-def} and \eqref{final-diagonalization-angles-def}:
    \begin{align}
    \label{heavy-scalars-limit-vanishing-angles}
    &\lim\limits_{\varepsilon\rightarrow 0}\tan{2\alpha} = \tan{2\beta} \Rightarrow \alpha = \beta,\quad \sin{2(\alpha - \beta)}\rightarrow 0,\quad \lim\limits_{\varepsilon\rightarrow 0}\alpha_3 = 0.
    \end{align}

    The self-interaction constants $\lambda^{(3)}$ and $\lambda^{(4)}$ then result in
    \begin{align*}
    \lim\limits_{\varepsilon\rightarrow 0}\lambda^{(3)} = -4v\lim\limits_{\varepsilon\rightarrow 0}\lambda^{(4)}.
    \end{align*}
    However, $m_h^2 \not\sim \lambda^{(4)}v^2$, as it takes place in the SM with one scalar doublet.

    Moreover, the transformation properties of $h$ are not identical to those of the SM Higgs boson, when $\varepsilon\rightarrow 0$. In this limit, $h$ does not become a CP-even field, as in the one-Higgs-doublet SM. Even when additional scalar bosons become heavy, the mixing angle $\alpha_2$ does not vanish, so $h$ contains contribution of CP-odd states $c_1$ and $c_2$, which is proportional to $\alpha_2$,
    \begin{align}
    \label{heavy-scalars-limit-non-vanishing-angle}
    \lim\limits_{\varepsilon\rightarrow 0}\alpha_2 = \frac{\Im\,\lambda_5 c_{2\beta}s_{2\beta}}{t_{2\beta}(\lambda_1 c_{\beta}^2 - \lambda_2 s_{\beta}^2) - 2s_{2\beta}\Re\,\lambda_5 - s_{2\beta}(\lambda_1 c_{\beta}^2 + \lambda_2 s_{\beta}^2)}.
    \end{align}
    Hence, some effects of CP violation could be detected in processes with the light $h$ boson.

    \section{Discussion and conclusions}
    \label{sec:conclusions}
    In the previous sections we discussed the scenario when one of the 2HDM scalar particles has the mass equalled  to that of the SM Higgs boson, and the other scalar states are heavy. We have obtained the analytical expressions for the effective operators describing interactions between the SM fermions and the lightest particle of the two-Higgs-doublet model, in terms of its parameters.

    We have shown also that the low-energy effective Lagrangian of the 2HDM in the decoupling limit does not transform to the Lagrangian of the one-Higgs-doublet model. Precise measurements of the triple and quartic self-couplings of the Higgs field could be used to discern the one-Higgs-doublet model from the 2HDM at low energies.

    The considered potential of the scalar fields also introduces a small CP-violation. It was shown that the angle $\alpha_2$, which describes mixing of scalars with opposite CP-parity, does not vanish in the limit when heavy scalars decouple, and the lightest neutral mass eigenstate of the model is not the eigenstate of the CP transformation. Hence, CP-violation in the 2HDM is potentially visible in modern experiments, and additional interactions within the scalar sector could be identified. Also, we found that the parameters $\alpha_2$ and $\alpha_3$  contribute the effective vertexes in the low-energy EL \eqref{effective-lagrangian}.

    At the tree level, the 2HDM introduces reactions mediated by charged scalars $H^{\pm}$, which are  absent in the SM with one Higgs doublet. In the low energy region, these processes are described by effective operators $J^{+(q)}J^{-(q)}$, $J^{+(l)}J^{-(l)}$, $J^{+(q)}J^{-(l)}$ and $J^{+(l)}J^{-(q)}$. Similar processes take place in the SM, too, but they are mediated only by the vector bosons $W^{\pm}$.

    Effective Lagrangian \eqref{effective-lagrangian} also introduces some new vertexes, which describe annihilation of a fermion-antifermion pair and the subsequent production of two or three Higgs bosons.

    Numerical predictions of the model with the EL \eqref{effective-lagrangian} are left beyond the scope of the present paper. They will  be studied in a separate publication.

    \section*{Appendix}
    The mass matrices of the scalar fields in 2HDM are
    \begin{align}
    \label{mass-matrices}
    M_a^2 &= \left[\frac{\Re m_{12}^2}{v_1 v_2} - \frac{1}{2}\left(\lambda_4 + \Re\lambda_5\right)\right]\begin{pmatrix}
    v_2^2 & -v_1 v_2 \\
    -v_1 v_2 & v_1^2
    \end{pmatrix}, \nonumber \\
    M_{bc}^2 &= \begin{pmatrix}
    \frac{v_2}{v_1}\Re\,m_{12}^2 + \lambda_1v_1^2 & -\Re\,m_{12}^2 + \lambda_{345}v_1 v_2 & \frac{v_2}{v_1}M_{bc23}^2 & -M_{bc23}^2 \\
    -\Re\,m_{12}^2 + \lambda_{345}v_1 v_2 & \frac{v_1}{v_2}\Re\,m_{12}^2 + \lambda_2 v_2^2 & M_{bc23}^2 & -\frac{v_1}{v_2}M_{bc23}^2 \\
    \frac{v_2}{v_1}M_{bc23}^2 & M_{bc23}^2 & M_{bc33}^2 & -\frac{v_1}{v_2}M_{bc33}^2 \\
    -M_{bc23}^2 & -\frac{v_1}{v_2}M_{bc23}^2 & -\frac{v_1}{v_2}M_{bc33}^2 & \frac{v_1^2}{v_2^2}M_{bc33}^2
    \end{pmatrix}, \nonumber \\
    M_{bc23}^2 &= \frac{1}{2}\Im\,\lambda_5 v_1v_2,\quad M_{bc33}^2 = \frac{v_2}{v_1}\Re\,m_{12}^2 - \Re\,\lambda_5 v_2^2.
    \end{align}
    Yukawa's interactions of the 2HDM mass eigenstates with the SM fermions is described by the terms in \eqref{yukawa-lagrangian-mass-eigenstates}. The contributions of quarks $J^{\pm(q)}$, $J_H^{(q)}$, $J_A^{(q)}$ and $J_h^{(q)}$ are as follows:
    \begin{equation}
    \label{quark-scalar-current-hp}
    J^{-(q)} = \sum\limits_{f;f^{\prime}}\left[\left(y_{ff^{\prime}}^{2(1)(q)}c_{\beta} - y_{ff^{\prime}}^{1(1)(q)}s_{\beta}\right)\overline{u}_L^{(f)}d_R^{(f^{\prime})} + \left(y_{ff^{\prime}}^{1(2)(q)*}s_{\beta} - y_{ff^{\prime}}^{2(2)(q)*}c_{\beta}\right)\overline{u}_R^{(f^{\prime})}d_L^{(f)}\right],
    \end{equation}
    \begin{align}
    \label{quark-scalar-current-h}
    J_H^{(q)} &= \frac{1}{\sqrt{2}}\sum\limits_{f;f^{\prime}}\left\{\left[-y_{ff^{\prime}}^{1(1)(q)}s_{\alpha} + y_{ff^{\prime}}^{2(1)(q)}c_{\alpha} + i\alpha_2\left(-y_{ff^{\prime}}^{1(1)(q)}s_{\beta} + y_{ff^{\prime}}^{2(1)(q)}c_{\beta}\right)\right]\overline{d}_L^{(f)}d_R^{(f^{\prime})} + \right. \nonumber \\
    &\left.+ \left[-y_{ff^{\prime}}^{1(2)(q)}s_{\alpha} + y_{ff^{\prime}}^{2(2)(q)}c_{\alpha} + i\alpha_2\left(y_{ff^{\prime}}^{1(2)(q)}s_{\beta} - y_{ff^{\prime}}^{2(2)(q)} c_{\beta}\right)\right]\overline{u}_L^{(f)}u_R^{(f^{\prime})} + h.\,c.\right\},
    \end{align}
    \begin{align}
    \label{quark-scalar-current-a0}
    J_A^{(q)} &= \frac{1}{\sqrt{2}}\sum\limits_{f;f^{\prime}}\left\{y_d\overline{d}_L^{(f)}d_R^{(f^{\prime})} + y_u\overline{u}_L^{(f)}u_R^{(f^{\prime})} + h.\,c.\right\}, \nonumber \\
    y_d &= i\left[-y_{ff^{\prime}}^{1(1)(q)}s_{\beta} + y_{ff^{\prime}}^{2(1)(q)}c_{\beta} + i\left(-y_{ff^{\prime}}^{1(1)(q)}(\alpha_3 c_{\alpha} + \alpha_2 s_{\alpha}) + y_{ff^{\prime}}^{2(1)(q)}(\alpha_2 c_{\alpha} - \alpha_3 s_{\alpha})\right)\right], \nonumber \\
    y_u &= i\left[y_{ff^{\prime}}^{1(2)(q)}s_{\beta} - y_{ff^{\prime}}^{2(2)(q)}c_{\beta} + i\left(-y_{ff^{\prime}}^{1(2)(q)}(\alpha_3 c_{\alpha} + \alpha_2 s_{\alpha}) + y_{ff^{\prime}}^{2(2)(q)}(\alpha_2 c_{\alpha} - \alpha_3 s_{\alpha})\right)\right],
    \end{align}
    \vspace{-5mm}
    \begin{align}
    \label{quark-scalar-current-light-h}
    J_h^{(q)} &= \frac{1}{\sqrt{2}}\sum\limits_{f;f^{\prime}}\left\{\left[-y_{ff^{\prime}}^{1(1)(q)}c_{\alpha} - y_{ff^{\prime}}^{2(1)(q)}s_{\alpha} + i\alpha_3\left(-y_{ff^{\prime}}^{1(1)(q)}s_{\beta} + y_{ff^{\prime}}^{2(1)(q)} c_{\beta}\right)\right]\overline{d}_L^{(f)}d_R^{(f^{\prime})} +\right. \nonumber \\
    &\left.+ \left[-y_{ff^{\prime}}^{1(2)(q)}c_{\alpha} - y_{ff^{\prime}}^{2(2)(q)}s_{\alpha} + i\alpha_3\left(y_{ff^{\prime}}^{1(2)(q)}s_{\beta} - y_{ff^{\prime}}^{2(2)(q)} c_{\beta}\right)\right]\overline{u}_L^{(f)}u_R^{(f^{\prime})} + h.\,c.\right\},
    \end{align}
    The contributions of leptons are analytically the same. They could be found if one substitutes $u$-type quarks with neutrinos and $d$-type quarks with electrons of the corresponding generation.

    From the Yukawa Lagrangian \eqref{yukawa-lagrangian} we also have the mass terms for the fermion  fields
    \begin{align}
    \label{fermionic-mass-terms}
    -\mathcal{L}_{mass} &= \frac{1}{\sqrt{2}}\sum\limits_{i=1,2}v_i\sum\limits_{f;f^{\prime}}\left[y_{ff^{\prime}}^{i(1)(q)}\overline{d}_L^{(f)}d_R^{(f^{\prime})} + y_{ff^{\prime}}^{i(2)(q)}\overline{u}_L^{(f)}u_R^{(f^{\prime})} +\right. \nonumber \\
    &\left.+ y_{ff^{\prime}}^{i(1)(l)}\overline{e}_L^{(f)}e_R^{(f^{\prime})} + y_{ff^{\prime}}^{i(2)(l)}\overline{\nu}_L^{(f)}\nu_R^{(f^{\prime})} + h.c.\right].
    \end{align}

\end{document}